\documentclass[11pt]{article}
\usepackage{graphicx}
\usepackage[margin=1.1in]{geometry}
\usepackage[usenames,dvipsnames]{color}
\usepackage{url}
\usepackage[colorlinks = true,
            linkcolor = blue,
            urlcolor  = blue,
            citecolor = blue,
            anchorcolor = blue]{hyperref}

\usepackage{epsfig}
\usepackage{cite}

\def\Title#1{\begin{center} {\LARGE #1 } \end{center}}
\def\Author#1{\begin{center}{ \sc #1} \end{center}}
\def\Address#1{\begin{center}{ \it #1} \end{center}}

\newenvironment{Abstract}{\begin{quotation} \begin{center}
                       ABSTRACT
     \end{center}\bigskip  }{\end{quotation}}

\newcommand{\mysection}{\setcounter{equation}{0}\section}

\def\beq{\begin{equation}}
\def\eeq{\end{equation}}
\def\beqa{\begin{eqnarray}}
\def\eeqa{\end{eqnarray}}

\newcommand\snowmass{\begin{center}\rule[-0.2in]{\hsize}{0.01in}\\\rule{\hsize}{0.01in}\\
\vskip 0.1in Submitted to the  Proceedings of the US Community Study\\ 
on the Future of Particle Physics (Snowmass 2021)\\ 
\rule{\hsize}{0.01in}\\\rule[+0.2in]{\hsize}{0.01in} \end{center}}

\begin{document}

\Title{Higher-order corrections for $t{\bar t}$ production at high energies}

\bigskip 

\Author{Nikolaos Kidonakis}

\medskip

\Address{Department of Physics, Kennesaw State University, \\
Kennesaw, GA 30144, USA}

\medskip

\begin{Abstract}
I present theoretical calculations of $t{\bar t}$ production cross sections through aN$^3$LO in proton-proton collisions. I show that soft-gluon corrections dominate the perturbative expansion not only for LHC energies but also for future collider energies through 100 TeV. Detailed results are presented for various collision energies. The $K$-factors are large and increase slowly with increasing energy. The scale uncertainties decrease greatly as we move to higher perturbative orders. 
\end{Abstract}

\snowmass

\def\thefootnote{\fnsymbol{footnote}}
\setcounter{footnote}{0}

\mysection{Introduction}

Top-quark production is a process of great interest in particle physics (see Ref. \cite{NKrev} for a review). The QCD corrections to $t{\bar t}$ production cross sections are very significant at NLO \cite{NLO1,NLO2} and NNLO \cite{NNLO}. The theoretical framework of soft-gluon resummation \cite{NKGS} provides powerful techniques for making theoretical predictions for QCD perturbative corrections at even higher orders. The resummation formalism for $t{\bar t}$ production developed in \cite{NKGS,NKloop} has produced very accurate predictions through approximate N$^3$LO (aN$^3$LO) \cite{NKN3LO} which is the current state of the art. The soft-gluon corrections are excellent approximations to the complete NLO and NNLO corrections at Tevatron and LHC energies (see e.g. Ref. \cite{NKrev}), and the additional aN$^3$LO corrections are significant and they provide an improved theoretical prediction. In this contribution, I present results for the cross sections through aN$^3$LO for various collider energies, and I show that the soft-gluon corrections are dominant even at very high energies through 100 TeV.

\mysection{Total cross sections for $t{\bar t}$ production}

In this section we present results for the cross section for $t{\bar t}$ production at LO, NLO \cite{NLO1,NLO2}, NNLO \cite{NNLO}, and aN$^3$LO \cite{NKN3LO}. Here we limit our study to total cross sections, but results for top-quark single-differential and double-differential distributions through aN$^3$LO can be found in Ref. \cite{NKN3LO}.

At leading order in the strong coupling for $t{\bar t}$ production, i.e. ${\cal O}(\alpha_s^2)$, the partonic channels are quark-antiquark annihilation, 
\beq
q(p_a)+{\bar q}(p_b) \to t(p_1) +{\bar t}(p_2) \, , 
\eeq
and gluon-gluon fusion,
\beq
g(p_a)+g(p_b) \to t(p_1) +{\bar t}(p_2) \, .
\eeq
We will use single-particle-inclusive kinematics in which the top quark with momentum $p_1$ is observed. If a soft-gluon with momentum $p_g$ is emitted, then the variable $s_4=(p_2+p_g)^2-m_t^2$ vanishes when $p_g$ goes to zero. Thus, this threshold variable describes the energy in the soft emission.

The resummation of soft gluons follows from the factorization properties of the cross section and the renormalization-group evolution of functions that describe the emission of soft and collinear gluons in the process \cite{NKGS,NKloop}. 
The $n$th order soft-gluon corrections to the (differential, in general) partonic 
cross section, $d{\hat \sigma}$, take the form 
\beq 
d{\hat \sigma}^{(n)}=\alpha_s^{n+2} 
\sum_{k=0}^{2n-1} C_k^{(n)}(s_4) \left[\frac{\ln^k(s_4/m_t^2)}{s_4}\right]_+ \, .
\eeq
The coefficients $C_k^{(n)}$ are in general functions of $s_4$ and other kinematical variables, the top-quark mass, the renormalization scale $\mu_R$, 
and the factorization scale $\mu_F$ \cite{NKGS,NKloop,NKN3LO}.

In addition to the usual terminology of LO, NLO, and NNLO, in the discussion below we will also employ the following terms: approximate NLO (aNLO), approximate NNLO (aNNLO), and approximate N$^3$LO (aN$^3$LO). The definition of aNLO is the following: aNLO = LO + soft-gluon NLO corrections; in other words, aNLO denotes the sum of the LO cross section and the NLO soft-gluon corrections. Hence, a comparison of aNLO and NLO results shows how dominant the soft-gluon corrections are. 

Similarly, aNNLO = NLO + soft-gluon NNLO corrections; in other words, aNNLO denotes the sum of the complete NLO cross section and the NNLO soft-gluon corrections, and a comparison of aNNLO and NNLO results again indicates the relative dominance of soft-gluon corrections. 

Finally, aN$^3$LO = NNLO + soft-gluon N$^3$LO corrections, i.e. aN$^3$LO is the sum of the complete NNLO corrections and the aN$^3$LO soft-gluon corrections, and it is the most precise result available for the $t{\bar t}$ cross section.

We use MSHT20 NNLO pdf \cite{MSHT20} for all our numerical results. Since we are interested in comparing the relative sizes of the contributions from each perturbative order to the total cross section, we use the same NNLO pdf for all results at all orders. We also set a common scale $\mu=\mu_F=\mu_R$ and choose $\mu=m_t=172.5$ GeV for our central results. The variation of $\mu$ from $m_t/2$ to $2m_t$ provides our result for the scale dependence of the cross section.

\begin{figure}[htbp]
\begin{center}
\includegraphics[width=81mm]{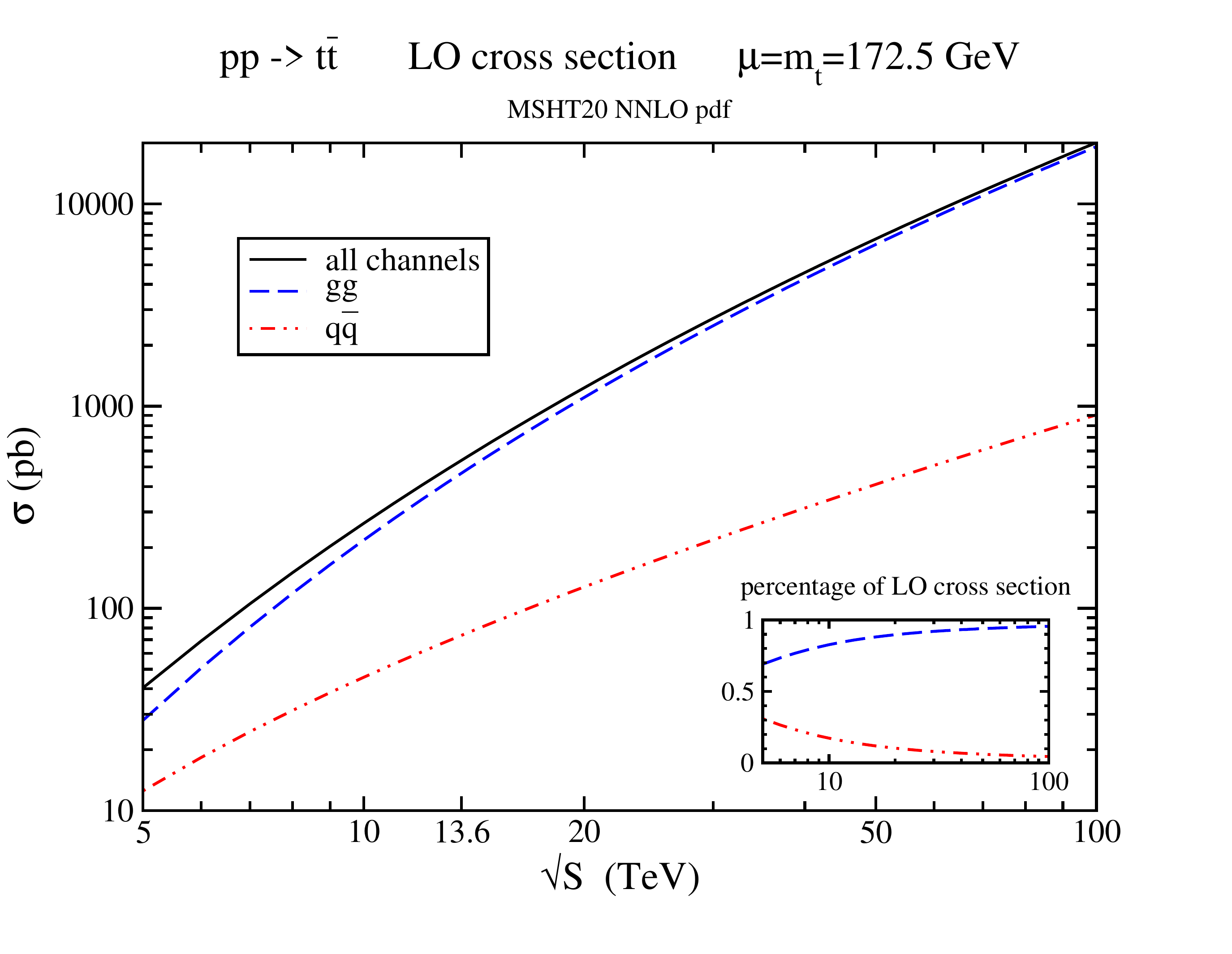}
\includegraphics[width=81mm]{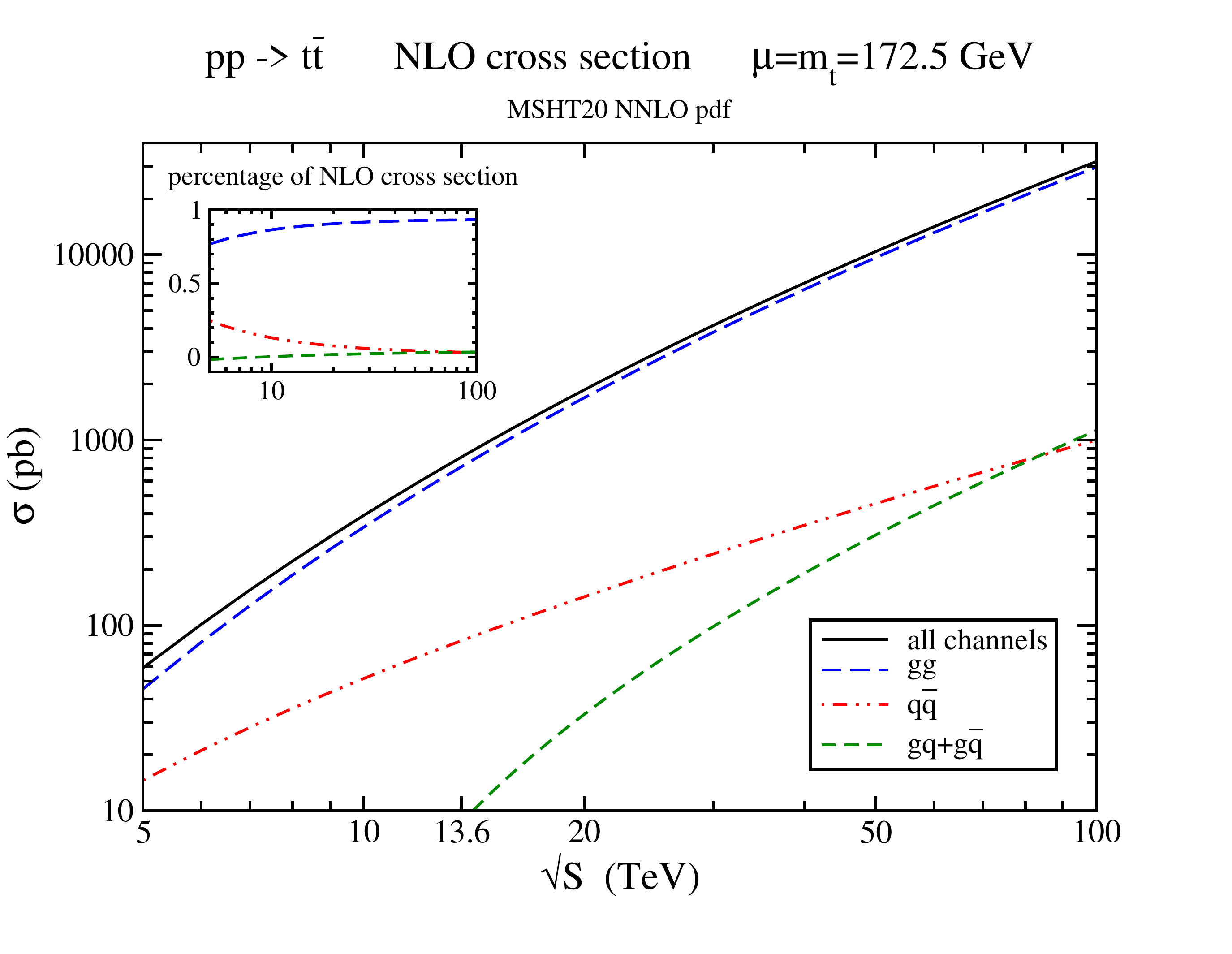}
\end{center}
\caption{The total cross sections at LO (left) and NLO (right) for $t{\bar t}$ production at $pp$ collider energies.}
\label{ttLONLOplot}
\end{figure}

We begin with LO and NLO results.
In the left plot of Fig. \ref{ttLONLOplot} we display the LO total cross sections with $\mu=m_t$ for $pp$ collider energies ranging from 5 TeV to 100 TeV. The inset plot displays the percentage contributions of the $gg$ and $q{\bar q}$ channels to the LO cross section. The plot on the right in Fig. \ref{ttLONLOplot} displays the total NLO cross sections, and the inset shows the percentage contributions of the $gg$, $q{\bar q}$, and $gq+g{\bar q}$ channels to the NLO cross section. While the cross section for each channel rises with energy, the percentage contribution to the total cross section, at both LO and NLO, rises in the $gg$ channel but it diminishes in the $q{\bar q}$ channel, reaching very low percentages at high energies. At NLO, there are also small contributions from the $gq$ and $g {\bar q}$ channels. For most of the energy range shown, the $gg$ channel accounts for over 90\% of the cross section at both LO and NLO. In fact, the $gg$ channel is overwhelmingly dominant in the entire range that is plotted.

\begin{table}[htb]
\begin{center}
\begin{tabular}{|c|c|c|c|c|c|c|c|c|c|} \hline
\multicolumn{10}{|c|}{$t{\bar t}$ cross sections in $pp$ collisions} \\ \hline
$\sigma$ in pb & \hspace{-3mm} 5.02 TeV \hspace{-3mm} & \hspace{-2mm} 7 TeV \hspace{-2mm} & \hspace{-2mm} 8 TeV \hspace{-2mm} & \hspace{-2mm} 13 TeV \hspace{-2mm} & \hspace{-3mm} 13.6 TeV \hspace{-3mm} & \hspace{-2mm} 14 TeV \hspace{-2mm} & 27 TeV & 50 TeV & 100 TeV \\ \hline
LO    & 41.0 & 106  & 150  & 488  & 540  & 576  & 2.23$\times 10^3$ & 6.72$\times 10^3$ & 20.1$\times 10^3$ \\ \hline
NLO   & 59.7 & 155  & 222  & 730  & 809  & 864  & 3.39$\times 10^3$ & 10.4$\times 10^3$ & 31.8$\times 10^3$\\ \hline
NNLO  & 67.1 & 174  & 249  & 814  & 902  & 963  & 3.77$\times 10^3$ & 11.5$\times 10^3$ & 35.1$\times 10^3$  \\ \hline
aN$^3$LO & 70.2 & 181 & 258 & 839 & 928 & 990 & 3.86$\times 10^3$ & 11.7$\times 10^3$  & 35.8$\times 10^3$ \\ \hline
\end{tabular}
\caption[]{The $t{\bar t}$ cross sections (in pb, with $\mu=m_t$) at different perturbative orders in $pp$ collisions with various values of $\sqrt{S}$, with $m_t=172.5$ GeV and MSHT20 NNLO pdf.}
\label{table1}
\end{center}
\end{table}

In Table 1 we show total cross sections for $t{\bar t}$ production for a variety of $pp$-collider energies, including LHC energies of 5.02, 7, 8, 13, 13.6, and 14 TeV as well as possible future-collider energies of 27, 50, and 100 TeV. We display the LO, NLO, NNLO, and aN$^3$LO cross sections at each energy. The results are calculated with a central choice of scale, $\mu=m_t$. We observe that the cross sections span three orders of magnitude as the energy ranges from 5 to 100 TeV.

\begin{figure}[htbp]
\begin{center}
\includegraphics[width=11cm]{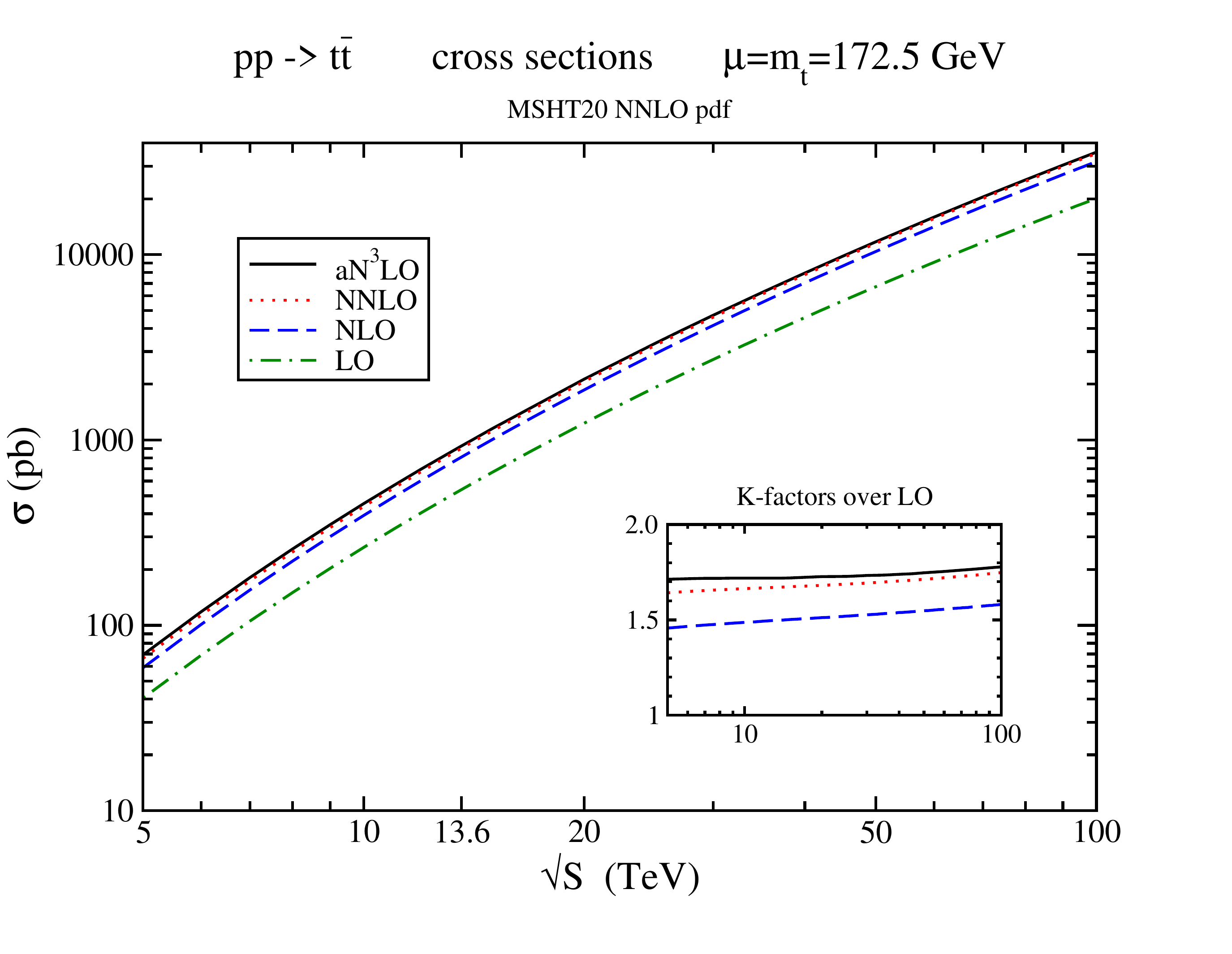}
\end{center}
\caption{The total cross sections at LO, NLO, NNLO, and aN$^3$LO for $t{\bar t}$ production at $pp$ collider energies.}
\label{ttaN3LOlogplot}
\end{figure}

In Fig. \ref{ttaN3LOlogplot} we show the LO, NLO, NNLO, and aN$^3$LO cross sections for $pp$ collider energies ranging from 5 TeV to 100 TeV. The inset plot displays the $K$-factors, i.e. the ratios of the NLO, NNLO, and aN$^3$LO cross sections to the LO ones. All three $K$-factors increase slowly with collision energy.

\begin{table}[htb]
\begin{center}
\begin{tabular}{|c|c|c|c|c|c|c|c|c|c|} \hline
\multicolumn{10}{|c|}{$K$-factors for $t{\bar t}$ production in $pp$ collisions} \\ \hline
$K$-factor & \hspace{-3mm} 5.02 TeV \hspace{-3mm} & \hspace{-2mm} 7 TeV \hspace{-2mm} & \hspace{-2mm} 8 TeV \hspace{-2mm} & \hspace{-2mm} 13 TeV \hspace{-2mm} & \hspace{-3mm} 13.6 TeV \hspace{-3mm} & \hspace{-2mm} 14 TeV \hspace{-2mm} & \hspace{-2mm} 27 TeV \hspace{-2mm} & \hspace{-2mm} 50 TeV \hspace{-2mm} & \hspace{-2mm} 100 TeV \hspace{-2mm} \\ \hline
NLO/LO  & 1.46 & 1.47 & 1.48 & 1.50 & 1.50 & 1.50 & 1.52 & 1.55 & 1.58 \\ \hline
NNLO/LO & 1.64 & 1.65 & 1.66 & 1.67 & 1.67 & 1.67 & 1.69 & 1.71 & 1.75 \\ \hline
aN$^3$LO/LO & 1.71 & 1.72 & 1.72 & 1.72 & 1.72 & 1.72 & 1.73 & 1.75 & 1.78 \\ \hline
aNLO/NLO & 1.02 & 1.01  & 1.00  & 0.99  & 0.99  & 0.99 & 0.97  & 0.95 & 0.92 \\ \hline
aNNLO/NNLO & 1.01 & 1.01 & 1.01 & 1.00 & 1.00 & 1.00 & 1.00 & 0.99 & 0.98 \\ \hline
\end{tabular}
\caption[]{The $K$-factors in $t{\bar t}$ production (with $\mu=m_t$) at different perturbative orders in $pp$ collisions with various values of $\sqrt{S}$, with $m_t=172.5$ GeV and MSHT20 NNLO pdf.}
\label{table2}
\end{center}
\end{table}

In Table 2 we show $K$-factors for $t{\bar t}$ production for the same $pp$-collider energies as in Table 1. The NLO/LO ratio is large for all energies, indicating large contributions from the NLO corrections. The NNLO/LO ratio is significantly larger, showing further important contributions from NNLO corrections. The aN$^3$LO /LO ratio is larger still, indicating further significant contributions from third-order soft-gluon corrections.

The dominance of the soft-gluon contributions for all energies is easily seen by the aNLO/NLO and the aNNLO/NNLO ratios, which remain very close to 1. Although the dominance of these corrections at LHC energies has been known for a long time (and reviewed in Ref. \cite{NKrev}), their continuing importance at very high energies is noteworthy and was not necessarily expected. Similar conclusions were drawn for the dominance of the soft-gluon corrections for $tW$ production through 100 TeV energy in Ref. \cite{NKNY}.

Next, we provide some cross sections with scale and pdf uncertainties at recent and future LHC energies. Again, we use MSHT20 NNLO pdf in all cases. The first set of uncertainties in each cross section given below is from scale variation while the second set shows the pdf uncertainties.

At 13 TeV, the LO cross section is $488^{+142}_{-103}{}^{+9}_{-6}$ pb, the NLO cross section is $730 \pm 87{}^{+15}_{-10}$ pb, the NNLO cross section is $814^{+28}_{-46}{}^{+16}_{-11}$ pb, and the aN$^3$LO cross section is $839^{+23}_{-18}{}^{+17}_{-11}$ pb.

At 13.6 TeV, the cross section is  $540^{+155}_{-113}{}^{+10}_{-7}$ pb at LO, $809^{+97}_{-95}{}^{+16}_{-11}$ pb at NLO, $902^{+31}_{-50}{}^{+18}_{-12}$ pb at NNLO, and $928^{+25}_{-20}{}^{+18}_{-12}$ pb at aN$^3$LO.

At 14 TeV, the cross section is  $576^{+164}_{-120}{}^{+11}_{-7}$ pb at LO, $864^{+103}_{-101}{}^{+17}_{-11}$ pb at NLO, $963^{+33}_{-53}{}^{+18}_{-13}$ pb at NNLO, and $990^{+27}_{-22}{}^{+19}_{-13}$ pb at aN$^3$LO.

We observe the increasing value of the central cross section with a decreasing scale dependence as we move up to higher orders, as expected, for all energies. The pdf uncertainties are consistently smaller than the scale uncertainties, even at aN$^3$LO.

\mysection{Conclusions}

I have presented results for $t{\bar t}$ production cross sections through aN$^3$LO at LHC energies as well as future collider energies up to 100 TeV. I have shown that the soft-gluon corrections dominate the cross section throughout the energy range studied from 5 TeV to 100 TeV. The cross section varies by three orders of magnitude over this energy range. The theoretical uncertainty due to scale dependence is very significantly reduced by the inclusion of higher-order corrections.

\section*{Acknowledgements}
This material is based upon work supported by the National Science Foundation under Grant No. PHY 2112025.

\end{document}